\documentclass[epj]{svjour}
\usepackage{latexsym}
\usepackage{graphics}
\usepackage{amsmath}
\usepackage{color}
\usepackage{graphicx}

\usepackage[english]{babel}
\usepackage{fancyref}
\usepackage[numbers,sort&compress]{natbib}
\usepackage[separate-uncertainty=true]{siunitx}
\usepackage{hyperref}
\hypersetup{
	colorlinks   = true,    
	urlcolor     = blue,    
	linkcolor    = blue,    
	citecolor    = blue     
}

\begin{document}

\title{Classical molecular dynamics simulations of fusion and fragmentation in fullerene--fullerene collisions}

\author{
Alexey Verkhovtsev\inst{1,2,}
\thanks{\email{verkhovtsev@iff.csic.es}}
\thanks{On leave from A. F. Ioffe Physical-Technical Institute, 194021 St. Petersburg, Russia}
\and
Andrei V. Korol\inst{1}
\and
Andrey V. Solov'yov\inst{1,}
\thanks{On leave from A. F. Ioffe Physical-Technical Institute, 194021 St. Petersburg, Russia}
}

\institute{
MBN Research Center, Altenh\"oferallee 3, 60438 Frankfurt am Main, Germany
\and
Instituto de F\'{\i}sica Fundamental, CSIC, Serrano 113-bis, 28006 Madrid, Spain
}


\abstract{
We present the results of classical molecular dynamics simulations of collision-induced fusion 
and fragmentation of C$_{60}$ fullerenes, performed by means of the MBN Explorer software package.
The simulations provide information on structural differences of the fused compound depending 
on kinematics of the collision process.
The analysis of fragmentation dynamics at different initial conditions shows that the size
distributions of produced molecular fragments are peaked for dimers, which is in agreement
with a well-established mechanism of C$_{60}$ fragmentation via preferential C$_2$ emission.
Atomic trajectories of the colliding particles are analyzed and different fragmentation
patterns are observed and discussed.
On the basis of the performed simulations, characteristic time of C$_2$ emission is estimated
as a function of collision energy.
The results are compared with experimental time-of-flight distributions of molecular
fragments and with earlier theoretical studies.
Considering the widely explored case study of C$_{60}$--C$_{60}$ collisions, we demonstrate
broad capabilities of the MBN Explorer software, which can be utilized for studying collisions
of a broad variety of nanoscale and biomolecular systems by means of classical molecular dynamics.
}

\authorrunning{Verkhovtsev, Korol, and Solov'yov}
\titlerunning{Classical MD simulations of fusion and fragmentation of fullerenes}

\maketitle

\section{Introduction}

Recent years have witnessed extensive development of experimental and theoretical methods
for the analysis of structure and dynamics of Meso-Bio-Nano (MBN) systems.
Irradiation and collision experiments have become a frequently used tool to explore
the internal structure and dynamical properties of such diverse systems.
As a result, a number of processes occurring in collisions of atoms, ions and
atomic clusters with complex molecular targets, biomolecules included, have been studied
\cite{ISACC_Latest_Advances2, ISACC2011_EPJD_Editorial, ISACC2015_EPJD_Editorial}.

Particular attention has been devoted to the investigation of irradiation- and collision-induced
processes involving carbon fullerenes, including
electron capture and ionization of C$_{60}$
\cite{Opitz_2000_PRA.62.022705, Tomita_2002_PRA.65.053201, Verkhovtsev_2013_PRA.88.043201};
fusion and fragmentation of C$_{60}$ induced by ion impact
\cite{Larsen_1999_EPJD.5.283, Manil_2003_PRL.91.215504, Rentenier_2008_PRL.100.183401,
Zettergren_2013_PRL.110.185501, Seitz_2013_JCP.139.034309}
or collisions with surfaces \cite{Fleischer_2011_NIMB.269.919, Bernstein_2016_JCP.145.044303};
fusion, fragmentation and charge transfer induced by fullerene--fullerene collisions
\cite{Campbell_1993_PRL.70.263, Rohmund_1996_PRL.76.3289, Rohmund_1996_JPB.29.5143,
Campbell_2000_RepProgPhys.63.1061, Brauning_2003_PRL.91.168301, Campbell_2003_book,
Jakowski_2012_JPCL.3.1536, Handt_2015_EPL.109.63001};
fission of C$_{60}$ irradiated with short intense laser pulses \cite{Fischer_2013_PRA.88.061403R},
and the formation of collective electron excitations due to photon
\cite{Hertel_1992_PRL.68.784, Scully_2005_PRL.94.065503, Baral_2016_PRA.93.033401},
electron \cite{Mikoushkin_1998_PRL.81.2707, Bolognesi_2012_EPJD.66.254, Schueler_2016_SciRep.6.24396}
and ion \cite{Kadhane_2003_PRL.90.093401, Kelkar_2010_PRA.82.043201} impact.

Recent advances in the understanding of ion/atom interactions with isolated polycyclic aromatic
hydrocarbons (PAHs), fullerenes and their clusters were discussed in a recent review
\cite{Gatchell_2016_JPB.49.162001}.
Apart from well-known statistical fragmentation of carbon systems leading to evaporation of
C$_2$ dimers, specific non-statistical fragmentation channels resulting in a prompt single-atom
atom knockout have been observed
\cite{Stockett_2014_PRA.89.032701, Delaunay_2015_JPCL.6.1536, Stockett_2014_PCCP.16.21980,
Stockett_2015_JPCL.6.4504, Gatchell_2015_PRA.92.050702R, Wolf_2016_EPJD.70.85}.

A considerable progress has also been achieved in experimental studies of collision-induced
processes involving biomolecular systems.
The latter targets were considered either in form of isolated biomolecules in the gas phase or
as clusters containing up to several tens of molecules.
The biomolecular targets have been characterized by an increasing complexity, starting from
water molecules and going to nucleobases, nucleosides and nucleotides, amino acids and
protein segments \cite{Nielsen_2004_JPB.37.R25, Schlatholter_2012_chapter, Rousseau_2017_chapter}.
Most of the experiments performed dealt with protons or multiply charged ions of carbon and oxygen,
i.e. the projectiles which are of current interest for ion-beam cancer therapy
\cite{Schardt_2010_RevModPhys.82.383, Surdutovich_2014_EPJD.68.353}.

The amount of accumulated experimental data on collision of atoms, ions and atomic clusters
with MBN systems generally exceeds the corresponding outcomes of theoretical and/or numerical analysis.
To a great extent, this disbalance can be attributed to the problems in finding efficient
theoretical and computational methods which allow one to accurately describe the collision-induced
dynamics in large molecular systems with many internal degrees of freedom.
It is therefore highly desirable to develop a single theoretical and computational tool
for modeling collision-induced processes involving different nanoscale systems.
This has become possible with the development of MBN Explorer software package
\cite{MBNExplorer_2012, MBNExplorer_book}.
The software supports the most advanced molecular dynamics (MD) simulations for a large
variety of complex molecular systems.
With these methods, one can simulate and study many different dynamical processes that occur
in molecular systems, including different collision and collision-induced processes.
By randomizing the initial conditions and carrying out multiple MD simulations, one can
generate sets of data for the statistical analysis of the outcome of the collision process.
This approach can be used, for example, for carrying out the analysis of mass-spectra of the
resulting fragments appearing in the course of collision.
Apart from the statistical analysis,
MD simulations allow one to visualize resulting atomic trajectories and explore the
temporal evolution of different molecular fragments.

A large number of force fields supported by MBN Explorer, together with its flexible and
efficient MD algorithms, allow one to model collision-induced dynamics of the ionic subsystem
of the colliding complexes of various types and internal structures, in broad range of collision
energies, and in various environmental and thermodynamical conditions.
The important general feature of collisions involving MBN systems arising from the fact
that they can be characterized not only by the collision energy, but also by temperature.
The colliding systems can be pre-equilibrated at a given temperature and then the
kinetic energy of the colliding objects can be fixed at some desirable value.
During the collision, part of the kinetic energy can be transferred to the internal degrees
of freedom of the colliding systems and be equilibrated there.
As a result, they may change their temperature, which may lead to the alteration of their
internal structure, as well as to evaporation, fragmentation and multi-fragmentation processes.


In this paper, we study collision-induced fusion and fragmentation of C$_{60}$ fullerenes
by means of classical MD simulations performed with MBN Explorer \cite{MBNExplorer_2012}.
C$_{60}$--C$_{60}$ collisions have been widely studied experimentally, and there are data on the
probability of fullerene fusion and on the production of smaller clusters due to subsequent
fragmentation \cite{Campbell_1993_PRL.70.263, Rohmund_1996_JPB.29.5143, Glotov_2000_PRA.62.033202}.
By considering this widely explored case study, 
we aim to demonstrate the main capabilities of the software that is
suitable for studying collisions of a broad variety of MBN systems.
The analysis of fragmentation dynamics shows that the size distributions of molecular fragments
produced are peaked for dimers, reflecting a well-established preferential C$_2$ emission.
Apart from that, the simulations provide information on structural aspects of the fused compound
at different kinematic conditions of the collision.
Finally, the atomic trajectories of the colliding particles are analyzed to explore the
dynamics of the collision events.
On the basis of the performed simulations, characteristic time of C$_2$ emission is estimated
as a function of collision energy.
The results are compared with experimental time-of-flight distributions of molecular fragments
and with earlier theoretical studies.

\section{Computational details}

MD simulations have been performed for the microcanonical ($NVE$) ensemble of particles,
where the number of particles $N$, the volume $V$, and the total energy $E$ of the system
were kept constant.
Integration of Newton's equations of motion was done using the velocity Verlet algorithm.
To assure conservation of the total energy, we used a small integration time step $\delta t = 0.1$\,fs.

The two fullerenes were placed in a $300 \times 300 \times 300$\,\AA$^3$ simulation box and were separated
by the distance of 50\,\AA~at the initial time moment.
The large size of the simulation box was chosen to decrease the probability of interaction between
small molecular fragments produced after the collision.
The simulations were performed for several collision energies and for different values of
the impact parameter.
The center-of-mass collision energy was varied from 30 eV
(corresponding to collision velocity $v = 40$~\AA/ps) up to about 370 eV ($v = 140$~\AA/ps).
We considered 15 values of the impact parameter,
ranging from 0\,\AA~(coaxial binary collision) up to 7\,\AA, which is approximately equal to
the diameter of C$_{60}$ (gliding collision).

In order to reflect the statistical nature of the fusion and fragmentation processes,
we performed 2000 simulations (80 simulations for a given collision energy
with different values of the impact parameter).
The simulation time for each run was 10\,ps that is of the same order of magnitude as
in many previous computational studies of fullerene fusion and fragmentation performed
by means of classical and quantum MD simulations
\cite{Bernstein_2016_JCP.145.044303,Zhang_1993_JPC.97.3134, Robertson_1995_JPC.99.15721,
Xia_1996_NIMB.111.41, Knospe_1996_JPB.29.5163, Glotov_2001_EPJD.16.333, Horvath_2008_PRB.66.075102}.
In each simulation run, the fullerenes were randomly oriented with respect to each other.
The input geometries were set up by means of MBN Studio \cite{MBNStudio_http, MBNStudio_tutorials} --
a graphical user interface to MBN Explorer.
The quantitative information on time evolution of the fragments produced (i.e., the number of
fragments of each type) has been obtained directly from the output of the simulations.
For each collision energy, ensemble-averaged fragment size distribution was calculated
by summing up the data from each individual trajectory and normalizing the resulting value
to the total number of fragments.

We employed the Brenner (reactive empirical bond-order, REBO) potential for carbon systems
\cite{Brenner_1990_PRB.42.9458}.
It is a many-body potential which depends on the number of nearest neighbors and contains
two-body and angle-depen\-dent three-body contributions.
The Brenner potential, al\-ongside with a similar many-body potential developed by Tersoff
\cite{Tersoff_1988_PRB.37.6991},
has been widely used for studying stability and
structural properties of many carbon systems, including fullerenes \cite{Robertson_1995_JPC.99.15721,
Yamaguchi_1998_CPL.286.336, Marcos_1999_EPJD.6.221, Chancey_2003_PRA.67.043203}
and nanotubes \cite{Lopez_2001_PRL.86.3056, Lopez_2005_Carbon.43.1371, Solovyov_2008_PRE.78.051601,
Verkhovtsev_2014_EPJD.68.246}.
Recently, these potentials were also utilized to study single and multiple atom knockouts from PAHs, 
fullerenes and their clusters (see Ref.~\cite{Gatchell_2016_JPB.49.162001} and references therein).

The total potential energy of a system in the Brenner potential framework reads as
\begin{equation}
U_{\rm tot} =
\frac12 \sum_{i} \sum_{i\ne j} f_{\rm c}(r_{ij}) \left[ U_{\rm R}(r_{ij}) - b_{ij} \, U_{\rm A}(r_{ij}) \right] \ ,
\label{eq:Tersoff2Basic}
\end{equation}
where $r_{ij}$ is the distance between atoms $i$ and $j$,
and $f_{\rm c}(r_{ij})$ is the cutoff function which limits the interaction of an atom to its nearest neighbors.
It is defined as
\begin{equation}
f_{\rm c}(r_{ij}) = \left\{
\begin{gathered}
1 \, , \ \ \ \ \ \ \ \ \ \ \ \ \ \ \ \ \ \ \ \ \ \ \ \ \ \ \ \ \ \ \ \ \ \ \ \ \ \ \ r_{ij} \leq R_{\rm 1}  \hfill \\
\frac12 + \frac12 \cos \left( \pi \frac{r_{ij}-R_{\rm 1}}{R_{\rm 2} - R_{\rm 1} } \right) \, , \ \ R_{\rm 1} < r_{ij} \leq R_{\rm 2} \hfill \\
0 \, , \ \ \ \ \ \ \ \ \ \ \ \ \ \ \ \ \ \ \ \ \ \ \ \ \ \ \ \ \ \ \ \ \ \ \ \ \ \ \ r_{ij} > R_{\rm 2}  \hfill \\
\end{gathered} \right.
\label{eq:cutoffTersoff2}
\end{equation}
with $R_{\rm 1,2}$ being the parameters which determine the range of the potential.
This function has a continuous value and derivative for all $r_{ij}$, and goes from 1 to 0
in a small region between $R_1$ and $R_2$,
which are chosen to restrict the potential to nearest neighbors.

The functions $U_{\rm R}(r_{ij})$ and $U_{\rm A}(r_{ij})$ are the repulsive and attractive
terms of the potential, respectively.
They are defined as
\begin{eqnarray}
U_{\rm R}(r_{ij}) &=& \frac{D_{\rm e}}  {S-1} \exp \left[ {-\sqrt{2S}\, \beta(r_{ij} - R_{\rm e})} \right] \nonumber \\
U_{\rm A}(r_{ij}) &=& \frac{D_{\rm e} \, S}{S-1} \exp \left[ {-\sqrt{\frac{2}{S}} \, \beta(r_{ij} - R_{\rm e})} \right] \ ,
\label{eq:VBrenner}
\end{eqnarray}
where $D_{\rm e}$, $S$, $\beta$ and $R_{\rm e}$ are parameters.

The factor $b_{ij}$ in Eq.~(\ref{eq:Tersoff2Basic}) is the so-called bond order term,
which is defined as:
\begin{equation}
b_{ij} = \left( 1 + \sum_{k\neq i,j} f_{\rm c}(r_{ik})g(\theta_{ijk}) \right)^{-\delta} \ .
\label{eq:BtermBrenner}
\end{equation}
Here, the function $g(\theta_{ijk})$ depends on the angle $\theta_{ijk}$ between bonds formed by pairs of atoms
$(i,j)$ and $(i,k)$.
This function has the following form:
\begin{equation}
g(\theta_{ijk}) = a \left[ 1 + \frac{c^2}{d^2} - \frac{c^2}{d^2 + (1 + \cos\theta_{ijk})^2} \right] \ ,
\end{equation}
where $a$, $c$ and $d$ are parameters of the potential.

The utilized parameters of the Brenner potential are listed in Table~\ref{table:parameters}.
\begin{table}[htb!]
\centering
\caption{Parameters of the Brenner \cite{Brenner_1990_PRB.42.9458}
potential used in the calculations.}
\begin{tabular}{p{1.2cm}p{2.0cm}p{1.2cm}p{2.0cm}}
\hline
$D_{\rm e}$ (eV)     &  6.325   & $a$                  &  0.011304  \\
$R_{\rm e}$ (\AA)    &  1.315   & $c$                  &  19    \\
$S$                  &  1.29    & $d$                  &  2.5   \\
$\beta$ (\AA$^{-1}$) &  1.5     & $R_{\rm 1}$ (\AA)    &  1.7    \\
$\delta$             &  0.80469 & $R_{\rm 2}$ (\AA)    &  2.0    \\
\hline
\end{tabular}
\label{table:parameters}
\end{table}

\section{Results and Discussion}
\label{Section_Results}

Let us now quantify the fusion and fragmentation products resulting
from the collision of two C$_{60}$ fullerenes at different collision velocities and
impact parameters.
In this work, collision products have been analyzed in the end of 10 ps-long simulations.
Fragmentation of fullerenes was also simulated on a few-picosecond timescale
in many previous studies employing classical and quantum MD approaches
(see, e.g., Refs. \cite{Zhang_1993_JPC.97.3134, Robertson_1995_JPC.99.15721, Xia_1996_NIMB.111.41,
Knospe_1996_JPB.29.5163, Glotov_2001_EPJD.16.333, Horvath_2008_PRB.66.075102}).
Most of these simulations were conducted for about $2-4$\,ps and demonstrated that the critical events
leading to fusion or multi-fragmentation of the colliding fullerenes occur very fast, within about 1\,ps.
The temporal evolution of different molecular fragments is analyzed in greater detail further in
this section.

\begin{figure}[t]
\centering
\resizebox{0.95\columnwidth}{!}{\includegraphics{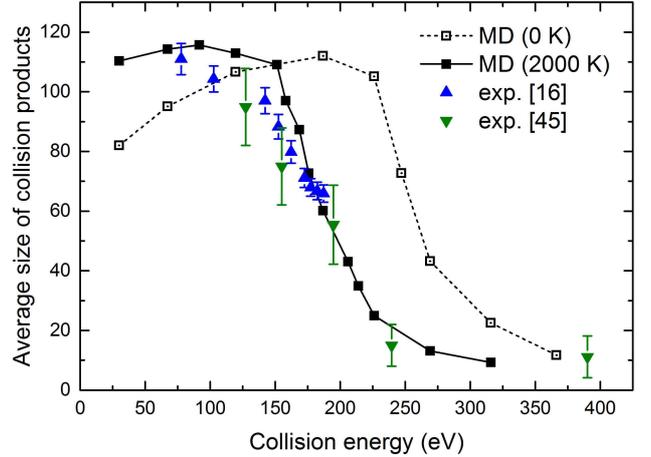}}
\caption{
The average size of molecular products produced in C$_{60}$-C$_{60}$ collisions as a function 
of the collision energy.
The collision products, including different molecular fragments as well as
non-fragmented C$_{60}$ molecules and fused C$_{120}$ compounds,
were recorded after 10\,ps of the simulations.
Open and filled squares describe the simulations performed at the fullerene initial
temperature of 0~K and 2000~K, respectively.
Other symbols represent experimental data from Refs.~\cite{Rohmund_1996_JPB.29.5143, Glotov_2000_PRA.62.033202}.
In the experiments, an average temperature of the colliding fullerenes was estimated around 2000~K.
}
\label{fig_avfragm}
\end{figure}

Figure~\ref{fig_avfragm} shows the average size of the molecular system recor\-ded at
the end of the simulations as a function of the center-of-mass collision energy.
The average system size was defined as the total number of atoms divided by the total
number of molecular species corresponding to a given collision energy.
As mentioned above, data extracted from many different trajectories at a given collision energy
were summed up and normalized to the total number of collision products, including different
molecular fragments as well as non-fragmented C$_{60}$ molecules and fused C$_{120}$ compounds.

\begin{figure}
\centering
\resizebox{0.95\columnwidth}{!}{\includegraphics{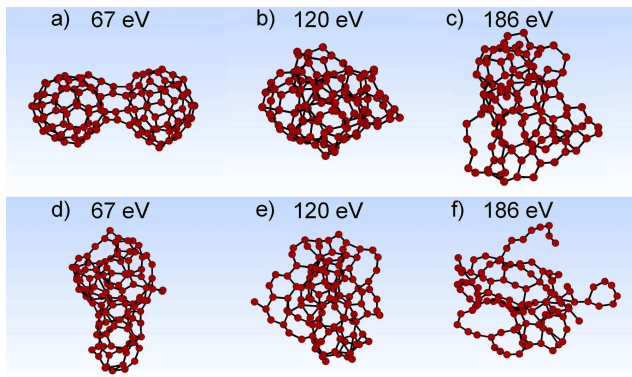}}
\caption{
Different isomers of C$_{120}$ formed after 5\,ps as a result of fusion of two
C$_{60}$ fullerenes at different collision energies.
The energies are indicated for each case study.
The upper row shows the structures which were simulated at zero fullerene temperature,
while the lower row corresponds to the fullerene temperature of 2000\,K.
The structures were rendered with MBN Studio software~\cite{MBNStudio_http}.
}
\label{fig_structures}
\end{figure}

Open squares in Figure~\ref{fig_avfragm} represent the results obtained at the zero
temperature of fullerenes.
Illustrative snapshots of the corresponding structures at different collision energies are
presented in the upper panel of Fig.~\ref{fig_structures}.
Figure~\ref{fig_avfragm} shows that the maximal average size of molecular products
and hence the maximal fusion probability
is obtained at collision energies of about 200\,eV, which is significantly higher than
experimental results obtained for ${\textrm C}_{60}^+ + {\textrm C}_{60}$ collisions
\cite{Rohmund_1996_JPB.29.5143, Glotov_2000_PRA.62.033202} (shown by blue and green triangles).
One should note that in the experiments, an average temperature of the colliding fullerenes
was estimated around 2000\,K \cite{Rohmund_1996_JPB.29.5143}.

In order to better match the experimental conditions, we performed simulations
where the fullerenes were given an initial temperature of 2000\,K.
As a result, each thermally excited molecule had an initial internal kinetic energy
of about 30\,eV.
Different initial structures and velocities used for the collision simulations were
obtained from a 10~ns-long constant-temperature simulation of a single C$_{60}$ molecule
being at $T=2000$\,K.
In this simulation, temperature control was achieved by means of the Langevin thermostat
with a damping time of 0.1~ps.
Note that similar simulations performed at different fullerene temperatures suggest that
C$_{60}$ resembles its intact cage-like structure up to $T \approx 2300$~K.
At higher temperatures, a transition, which is usually considered as the fullerene melting
takes place.
It corresponds to an opening of the fullerene cage and the formation of a highly-distorted
but still non-fragmented structure \cite{Kim_1994_PRL.72.2418, Hussien_2010_EPJD.57.207}.

The results of the simulations at $T = 2000$~K are shown in Fig.~\ref{fig_avfragm} by filled squares.
In agreement to what is known in the literature \cite{Robertson_1995_JPC.99.15721,
Zhang_1993_JPC.97.3134, Knospe_1996_JPB.29.5163},
a non-zero initial temperature of the fullerenes gives a much better agreement with the
experimental results.
Taking into account that statistical uncertainty of the calculated average size of collision products
is about 10\%, the calculated numbers agree well with the experimental data.
We found that the largest average product size and hence the highest probability of fusion
is for collisions with energies of about 90-120\,eV, which is significantly lower than the value 
of about 200~eV simulated at zero initial temperature.
The fusion barrier decreases due to the thermal energy stored in the fullerenes.

Earlier works \cite{Campbell_2000_RepProgPhys.63.1061, Glotov_2001_EPJD.16.333}
concluded that classical MD simulations usually provide much larger values for the energy
window for the fusion reaction as compared to the energy window observed experimentally.
In the analysis presented above, we demonstrated that classical MD is an adequate approach
which can reflect the main features of the collision-induced processes if the initial internal
energy of the projectile and the target is taken into account.
Some disagreement between the simulation results and experimental data can be attributed
to the way how the initial temperature was assigned to each fullerene.
In experiments \cite{Rohmund_1996_PRL.76.3289, Rohmund_1996_JPB.29.5143}, the target
fullerene was heated up to about 800\,K in the scattering cell,
while the temperature of the projectile was estimated to be as high as 3000\,K
\cite{Campbell_2000_RepProgPhys.63.1061, Rohmund_1996_JPB.29.5143}.
Accounting for the different initial temperatures of the projectile and the target may
improve the agreement between the theoretical results and experiment even further.

The above-presented analysis was performed for the collision of two neutral C$_{60}$ fullerenes,
while collisions between a singly-charged and a neutral system, C$_{60}^+$--C$_{60}$,
were studied experimentally \cite{Rohmund_1996_JPB.29.5143, Glotov_2000_PRA.62.033202}.
To explore the effect of an excess charge on the collision dynamics, we performed simulations of
C$_{60}^+$--C$_{60}$ collisions for selected collision energies (91, 151 and 186\,eV) at the initial
fullerene temperature of 2000\,K.
The positive charge was uniformly distributed over the projectile, so that each carbon atom
carried a partial charge of $+0.01667e$.
In the new set of simulations, we have not observed any statistical difference from the results
obtained for the two neutral molecules.
These results suggest that the effect of including charge in the simulations is very small and
can thus be neglected.
Charge effects may have a stronger impact on the collision dynamics in the case when one of the
colliding molecules has a higher charge state or both molecules are charged
\cite{Kamalou_2006_IJMS.252.117, Jakowski_2012_JPCL.3.1536}.
This is an interesting question that can be addressed in detail in further studies.

One should note that in the simulations of the C$_{60}^+$--C$_{60}$ collisions,
minor effects of charge redistribution have been observed.
Despite this, all the small fragments which have been produced were electrically neutral,
with either zero or small partial charge on different atoms.
The effects of charge redistribution can be elaborated in greater detail by means
of irradiation-driven molecular dynamics (IDMD), that is a novel approach for modeling
irradiation or collision-driven chemical transformations of complex molecular systems
\cite{Sushko_2016_EPJD.70.217}.
However, this is a separate scientific problem which we do not aim to consider in this work.

Figure~\ref{fig_avfragm} demonstrates that at low collision energies (below about 100~eV),
the average collision product size decreases.
It happens because of the increasing probability of non-reactive inelastic scattering of
two fullerenes, which does not lead to fusion.
The complete fusion of two C$_{60}$ into one C$_{120}$ structure was observed at $T = 0$~K at
the energy of 120~eV (see Figure \ref{fig_structures}b).
This is in agreement with the results of density-functional tight-binding (DFTB) MD simulations
\cite{Jakowski_2010_PRB.82.125443} which showed that the energies higher than 100~eV are
needed to form a single-cage C$_{120}$.
In this energy region, the fusion process results in the formation of elongated peanut-shaped structures.

As known from the earlier theoretical studies \cite{Jakowski_2010_PRB.82.125443, Wang_2014_PRA.89.062708},
at lower energies, the two molecules tend to form a covalently bonded dimer (C$_{60})_2$,
and this process occurs when the collision energy is not high enough to break more than
one or a few bonds.
DFTB MD simulations \cite{Wang_2014_PRA.89.062708} predicted that the threshold collision energy
for this process is about 60~eV.
One should note, however, that we observed a significant probability of forming a covalently bonded
dimer even at the energy of 30~eV.
This observation corresponds to the results of earlier classical MD simulations using the Tersoff potential,
which gave higher formation probabilities of covalently bonded (C$_{60})_2$ dimers than within
the DFTB method (see the Supplemental Material in \cite{Wang_2014_PRA.89.062708}),
and predicted the kinetic energy threshold as low as 15~eV.
This feature was attributed to the fact that the Tersoff potential overestimates the bonding
between $sp^2$ and $sp^3$ carbon atoms.
The Brenner potential, which we have employed in this work, may have a similar deficiency.

In Figure \ref{fig_structures}, we compare the structure of a C$_{120}$ compound formed
as a result of the fullerene fusion.
The structures are shown for three collision energies, namely 67~eV (panels (a,d)), 120~eV (b,e)
and 186~eV (c,f), for both 0~K and 2000~K initial temperature of the colliding fullerenes
(upper and lower panels, respectively).
As discussed above, simulations performed at zero temperature and low collision energy result
in the formation of a dumbbell structure~(a).
The simulations at the same energy but at a finite fullerene temperature result
in the formation of a closed-cage peanut-like structure~(d).
A similar structure was obtained as a result of simulations performed at 0~K and 120~eV energy~(b).
This compound is highly deformed but still represents a closed-cage structure.
On the contrary, in the simulations at 2000~K and 120~eV, the energy deposited into the system
is enough to break the fullerene cage~(e).
This structure resembles the ``pretzel phase'', observed in earlier MD simulations of the
C$_{60}$ melting at $T \approx 4000$~K \cite{Kim_1994_PRL.72.2418}.
A similar open-cage structure was produced in 186~eV collision at zero fullerene temperature~(c).
At this collision energy, the presence of thermal energy of the fullerenes leads
to formation of a loosely bound structure with several linear chains (f), and this structure is
then subject to fragment.
This structure is similar to the ``linked chain'' phase, observed in the process of C$_{60}$
melting at temperatures above 5000~K \cite{Kim_1994_PRL.72.2418}.

\begin{figure}
\centering
\resizebox{0.95\columnwidth}{!}{\includegraphics{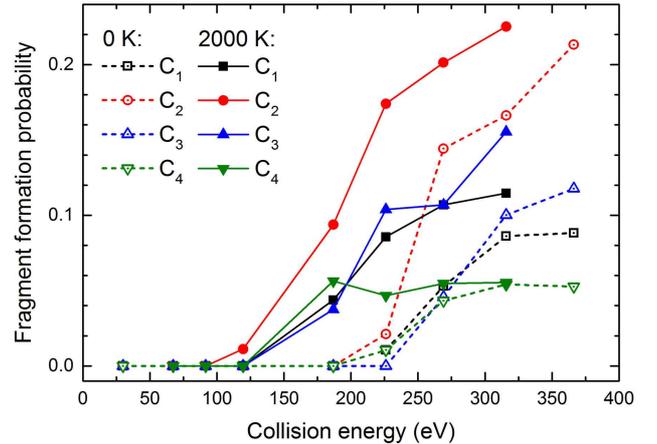}}
\caption{
Comparison of the probabilities of small fragments (C$_1$ -- C$_4$) formation
in the C$_{60}$--C$_{60}$ collisions as a function of the collision energy.
The fragments were recorded after 10\,ps of the simulations.
Open and filled symbols describe the simulations performed at the fullerene initial
temperature of 0\,K and 2000\,K, respectively.
}
\label{fig_fragprob}
\end{figure}

It is now commonly accepted that an abrupt decrease of the fusion signal, observed experimentally
at the collision energies around 200~eV, is an indication of the rapid loss of the fullerene
structure and the onset of a multi-fragmentation regime, leading to the production of many
small-size fragments \cite{Rohmund_1996_PRL.76.3289, Rohmund_1996_JPB.29.5143}.
In order to describe the multi-fragmentation process in more detail,
we have analyzed the formation of small fragments (C$_1$, C$_2$, C$_3$, C$_4$)
as a function of the center-of-mass collision energy.
The corresponding probabilities are shown in Figure \ref{fig_fragprob}.
These probabilities were defined as a ratio of the number of C$_1$ -- C$_4$ fragments formed
after 10~ps, to the total number of fragments produced.
Open symbols describe the results of simulations performed at the zero
initial temperature of colliding fullerenes, while filled symbols describe
the case of $T = 2000$~K.
It is seen that the formation probabilities show different trends for different fragments.
The probability of the dimer formation rapidly increases in both cases,
confirming that C$_2$ emission is the leading statistical channel of fullerene
fragmentation at moderate collision energies.
The probabilities for a single carbon atom and a tetramer formation gradually
saturate with increasing the collision energy.
However, the probability for C$_3$ formation also increases with energy,
especially in the case of nonzero fullerene temperature simulations.
This observation correlates with the results of earlier TB-MD simulations
\cite{Horvath_2008_PRB.66.075102} which studied radiation-induced fragmentation of C$_{60}$.
In the cited work, it was shown that C$_3$ becomes the most
probable pathway of the C$_{60}$ fragmentation at increasing excitation energy.

\begin{figure}
\centering
\resizebox{0.97\columnwidth}{!}{\includegraphics{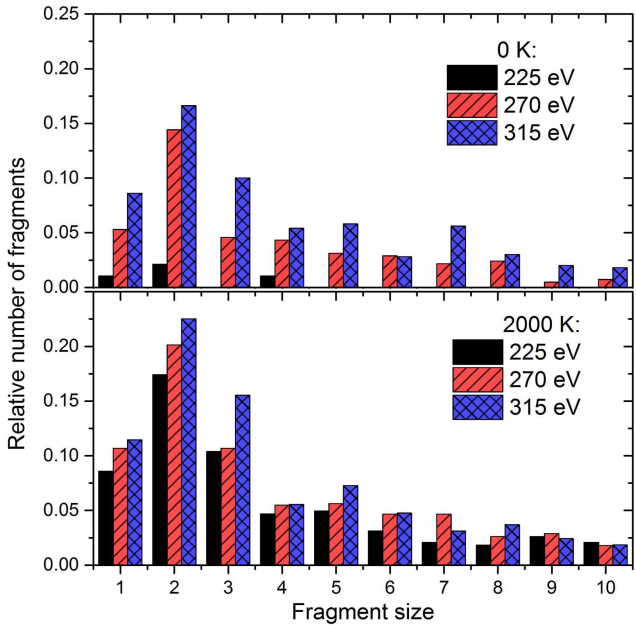}}
\caption{
Number of C$_1$ -- C$_{10}$ fragments, normalized to the total number of fragments
produced after 10~ps, for the center-of-mass collision energies of 225, 270 and 315 eV.
The upper and the lower panels show the results obtained at the 0~K and 2000~K
temperature of colliding fullerenes, respectively.
}
\label{fig_numberrfrag}
\end{figure}

To analyze in more detail the impact of the fullerene initial temperature on the fragmentation
dynamics, we plotted the size distribution of larger fragments, up to C$_{10}$,
formed in the end of 10~ps-long simulations.
The results of this analysis are shown in Figure~\ref{fig_numberrfrag}.
The simulations performed at zero initial temperature of fullerenes (upper panel) show
that at the collision energy of 225~eV, only a few fragmentation events have been observed,
while at the energy of 270~eV a phase transition has taken place leading to multi-fragmentation
of the fullerenes and the formation of multiple small-size fragments.
The results of simulations at $T=2000$~K fullerene temperature (lower panel)
demonstrate that the phase transition takes place at lower collision energy.
Our analysis shows that in this case, the multi-fragmentation regime starts at the
collision energy of about 185~eV.
As discussed above, the most prominent effect of the fullerene finite temperature
is an increase in the number of C$_2$ and C$_3$ fragments.
The data shown in Fig.~\ref{fig_numberrfrag} demonstrate that at 315~eV collision energy
the relative number of larger fragments is about 3-6\% of the total number of fragments produced,
and these values are almost independent on the initial energy stored in the system.

It is known that the size distribution of small fragments C$_n$, produced in collisions involving
fullerene mole\-cules, follows a $n^{-\lambda}$ power law
\cite{Rentenier_2005_JPB.38.789, Horvath_2008_PRB.66.075102}.
Having taken into account that the simulated distributions of fragments are peaked at $n=2$,
we have fitted the results for $n \ge 2$ with a power function.
As a result of the fitting procedure, we obtained the value of $\lambda = 1.47 \pm 0.04$, which is
close to the value of 1.54, obtained in earlier MD statistical trajectory simulations at 500~eV
center-of-mass collision energy \cite{Schulte_1995_IJMS.145.203}.

\begin{figure}
\centering
\resizebox{0.97\columnwidth}{!}{\includegraphics{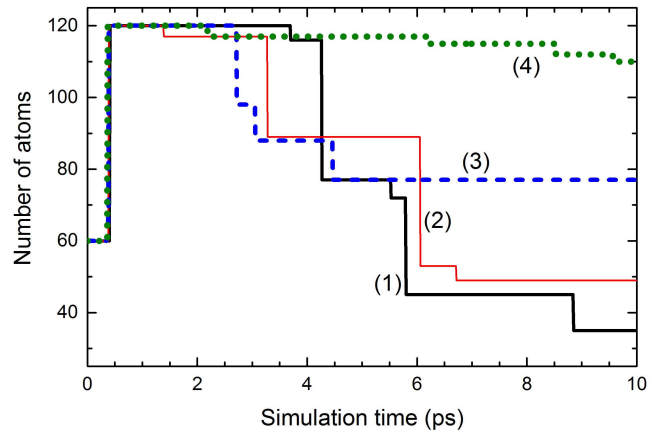}}
\caption{
Time evolution of the size of the largest molecular product during a 10-ps simulation.
Four representative atomic trajectories are shown by different lines.
}
\label{fig_trajectories}
\end{figure}

Apart from the statistical analysis of molecular fragments produced in the collisions,
MD simulations provide a possibility to visualize resulting atomic trajectories and explore
the temporal evolution of different molecular fragments.
To illustrate this, we analyzed four representative trajectories obtained at
226~eV center-of-mass collision energy at 2000~K.
Figure~\ref{fig_trajectories} shows how the size of the largest molecular product has been
evolving in the course of simulation.
The two colliding fullerenes have fused into a single compound after the first 0.4~ps
of the simulations as illustrated by a sharp jump in the number of atoms comprising the
largest product from 60 to 120.
However, the subsequent evolution of this system is quite different in the four considered
trajectories:
the lifetime of the fused compound varies between about 1 and 3.3~ps and the fragmentation channels
are also rather different.
For instance, in trajectory~1 (solid black curve) a C$_4$ tetramer was emitted first at about 3.7~ps,
and the resulting C$_{116}$ molecule dissociated into two large fragments containing 77 and 39 atoms.
The former fragment emitted a small carbon molecule and then also disintegrated into two large
products formed by 45 and 27 atoms.
After another fragmentation event, the largest molecule recorded after 10~ps of the simulation
has only 33 atoms.
On the contrary, trajectory 4 (dotted green curve) has been evolving in a completely different way:
no fragmentation into large products has been observed but the fused C$_{120}$ compound
sequentially emitted two dimers and two trimers, so that the final structure recorded after 10~ps
consists of 110~atoms.

\begin{figure}
\centering
\resizebox{0.97\columnwidth}{!}{\includegraphics{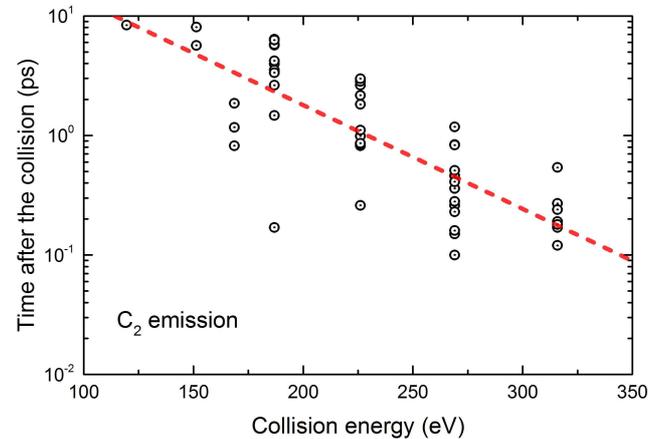}}
\caption{
Characteristic time of emission of C$_2$ fragments at different collision energies at 2000~K.
This quantity was defined as a lifetime of the fused C$_{120}$ compound before fragmenting
into $\rm{C}_{118} + \rm{C}_2$ or $\rm{C}_{118 - x} + \rm{C}_2 + \rm{C}_x$ products.
The data extracted from the simulations are shown by symbols, while the dashed line
shows a linear least-squares fit.
}
\label{fig_C2_emission_time}
\end{figure}

The information stored in the atomic trajectories can be used to explore the dynamics
of the collision events.
In particular, one can analyze characteristic times of emission of fragments of a
given size.
We have monitored emission of the most frequently produced fragmentation products,
C$_2$ dimers, at different collision energies at 2000~K;
the results of this analysis are shown in Figure~\ref{fig_C2_emission_time}.
The emission time was defined as a lifetime of the fused C$_{120}$ compound before
fragmentation into $\rm{C}_{118} + \rm{C}_2$ or $\rm{C}_{118 - x} + \rm{C}_2 + \rm{C}_x$ products.
One can see that at collision energies of about $100-150$~eV, i.e., before the multi-fragmentation
takes place, C$_2$ fragments are produced in small numbers and mostly within a time window of
$5-10$~ps after the two fullerenes had collided.
With an increase of the collision energy, the dimers start to eject from the system much faster.
At the center-of-mass collision energy of 315~eV, C$_2$ fragments are produced in much larger
numbers on a sub-picosecond time scale, thus indicating the multifragmentation regime.
One can expect that with a further increase of collision energy, the fragments will be produced
even faster, on the order of several tens of femtoseconds.
Note that similar behavior was also observed for other abundantly produced fragments like single
carbon atoms and C$_3$ molecules.

\section{Conclusion and Outlook}

This work has been devoted to the investigation of C$_{60}$ fullerene collisions
and collision-induced fusion and fragmentation processes by means of classical molecular
dynamics simulations performed with the MBN Explorer software package.
The simulations were performed in a broad range of collision energies, thus allowing to model
the formation of covalently-bonded dumbbell structures, closed-cage C$_{120}$ compounds,
open-cage structures, as well as sequential emission of small-size molecular fragments
and rapid multi-fragmentation.
We demonstrated that classical molecular dynamics is capable of describing the main features
of the collision-induced processes if the initial internal energy of the projectile is taken
into account.

We analyzed the fragmentation dynamics and showed that the size distributions of
molecular fragments produced are peaked for dimers, reflecting a well-known statistical
channel of C$_{60}$ fragmentation via emission of carbon dimers.
The performed atomistic simulations provided information on structural aspects of the
fused compound at different collision energies and thermal energy of the colliding
molecules.
Our results have been compared with well-established experimental results on
time-of-flight distributions of molecular fragments.
The simulation results have been found in agreement with the
experimental data and the results of earlier theoretical studies.
We demonstrated that, apart from statistical analysis of produced fragments, molecular dynamics
simulations performed with MBN Explorer allow one to analyze temporal evolution of these fragments.
In this work, we studied the temporal evolution of several atomic trajectories and evaluated
the characteristic time of emission of the most abundantly produced fragment, the C$_2$ dimer.

Performing this analysis, we presented some of the capabilities of MBN Explorer
to model collisional processes involving a broad range of Meso-Bio-Nano systems.
Although it is not possible to cover many different case studies in a single paper,
we note that by means of this tool, it is possible to model collision-induced processes
with many different nano- and biological systems.
A broad variety of interatomic potentials, including many-body potentials for
multicomponent systems, and the CHARMM molecular mechanics potential for organic
and biomolecular systems, are implemented in the software, allowing for all-atom
modeling of composite materials and nano-bio interfaces.
MBN Explorer provides also the tools to the multiscale modeling of collisions in which
the dynamics of Meso-Bio-Nano systems is accompanied by the random, local quantum
transformations of the system (such as ionization or charge transfer)
induced in the system during the collision process.
Recently, such possibilities have been implemented through the reactive force field
\cite{Sushko_2016_EPJD.70.12} and the irradiation-driven molecular dynamics
approach \cite{Sushko_2016_EPJD.70.217}.
The latter represents a classical molecular dynamics with the superimposed random processes
of local quantum transformations related to the irradiation conditions.
All these features allow modeling of the collision phenomena involving a broad variety of
nanoscale and biomolecular systems, including
such widely studied systems like PAHs,
novel materials like boron-nitride fullerenes and nanotubes,
metallic nanoalloys,
collisions with surfaces,
and many more.
A detailed analysis of the processes occurring in these systems is of great current interest
but goes well beyond the scope of a single paper.
Therefore, this analysis will be continued in future works.

\section*{Acknowledgements}

AV acknowledges the support by the European Commission through the FP7 Initial Training Network
``ARGENT'' (grant agreement no. 608163).

\section*{Author contribution statement}

AV performed the calculations, analyzed the results and drafted the manuscript.
All authors participated in the discussion of the results, provided valuable comments,
and contributed to the revision of the manuscript.


\end{document}